%% file: main.tex
\documentclass[conference]{IEEEtran}
\usepackage{cite}
\usepackage{amsmath,amssymb,amsfonts}
\usepackage{algorithmic}
\usepackage{graphicx}
\usepackage{textcomp}
\usepackage{xcolor}
\usepackage{hyperref}
\usepackage{siunitx}
\usepackage{svg}

\usepackage{lipsum}

\DeclareSIUnit\sop{SOP}
\DeclareSIUnit\inference{inference}
\DeclareSIUnit\GE{GE}
\DeclareSIUnit\pjsop{\pico\joule\per\sop}
\DeclareSIUnit\njinf{\nano\joule\per\inference}
\DeclareSIUnit\ujinf{\micro\joule\per\inference}
\DeclareSIUnit\GSOPs{\giga\sop\per\second}
\DeclareSIUnit\GSOPs{\tera\sop\per\second}

\def\BibTeX{{\rm B\kern-.05em{\sc i\kern-.025em b}\kern-.08em
    T\kern-.1667em\lower.7ex\hbox{E}\kern-.125emX}}

\begin{document}
\bstctlcite{IEEEexample:BSTcontrol}

\title{Mega: A 22 nm Convolutional Spiking Neural Network Accelerator Achieving 0.375 pJ/SOP for Efficient Edge Vision}

\author{\IEEEauthorblockN{Rick Luiken$^\star$, Manil Dev Gomony and Sander Stuijk}
\IEEEauthorblockA{Dept. of Electrical Engineering, Eindhoven University of Technology, The Netherlands \\
\textit{$^\star$Correspondence to h.j.luiken@tue.nl}
}}

\maketitle

\begin{abstract}
Convolutional Spiking Neural Networks (SNN) offer the potential for highly energy-efficient vision processing by exploiting sparse, event-driven computation. However, existing SNN accelerators underutilize the inherent parallelism of convolutional layers and lack the flexibility to accommodate varying memory demands and input sparsity across layers. This paper presents Mega, a digital architecture for convolutional SNNs that addresses these limitations through three key contributions: (1) highly parallel acceleration of $3 \times 3$ convolutions, (2) a unified data memory for spikes, neuron states, and weights, and (3) efficient spike map processing with low-overhead spike detection. Fabricated in GlobalFoundries \qty{22}{nm} FDSOI technology, Mega achieves an energy efficiency of \qty{0.375}{\pjsop}, improving the state of the art by $4\times$.
\end{abstract}


\section{Introduction}
Enabling vision-based applications at the edge, where energy and compute resources are scarce, requires highly efficient processing. Dynamic Vision Sensors (DVS) \cite{lichtsteiner2008} address this need by converting changes in visual scenes into sparse, event-driven spikes with high temporal resolution. Convolutional Spiking Neural Networks (SNNs) naturally exploit this sparsity, enabling energy-efficient inference for vision tasks.

Despite the potential, accelerating convolutional SNNs in hardware remains challenging. Existing SNN accelerators often fail to fully exploit the parallelism of convolutional layers, particularly for small kernels such as $3 \times 3$, which dominate network designs. Memory demands also vary across layers: initial layers often require high-resolution feature maps, making neuron state memory dominant, while deeper layers have many channels, making weight memory dominant. Additionally, common spike communication schemes, such as Address-Event Representation (AER), are efficient only at very high sparsity. However, typical sparsity levels vary per layer, reducing their effectiveness.

\begin{figure}
    \centering
    \includegraphics[width=\linewidth,trim=0 2 2 3,clip]{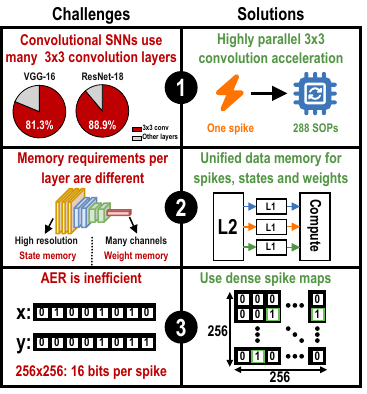}
    \caption{\textbf{Challenges and Solutions}. The left side illustrates common challenges encountered when accelerating convolutional SNNs, while the right side presents our corresponding solutions.}
    \label{fig:challenges}
\end{figure}

To address these challenges, this paper presents Mega, a digital architecture for convolutional SNNs. Mega introduces three key innovations, illustrated in \autoref{fig:challenges}:
\begin{enumerate}
    \item Highly parallel $3\times3$ spiking convolution acceleration, exploiting kernel-level and output channel parallelism.
    \item A unified data memory for spikes, neuron states, and weights that supports different layer types.
    \item Low-overhead spike detection and address conversion for efficient event-driven processing.
\end{enumerate}

\section{Hardware architecture}

\begin{figure*}[t]
\centering
\includegraphics[width=0.8\linewidth]{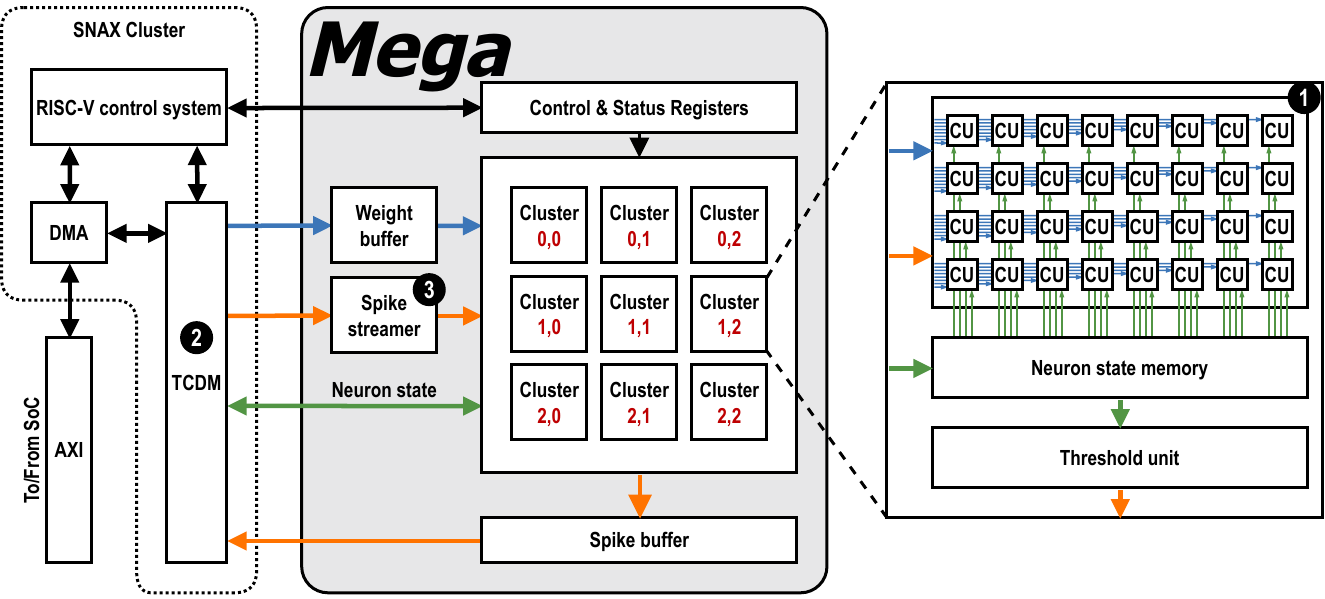}
\caption{\textbf{Mega architecture diagram}. Mega is built around nine compute clusters that compute $3 \times 3$ convolutions in parallel. Each cluster updates 32 neuron states in parallel using 32 convolution units (CUs). Spikes, weights, and neuron states are fetched from the integrated SNAX cluster \cite{antonio2025}. The spike streamer converts dense spike maps into sparse spike addresses, which are forwarded to the compute clusters.}
\label{fig:architecture}
\end{figure*}

\autoref{fig:architecture} shows the hardware architecture of Mega. The design is centered around nine compute clusters that collectively perform parallel 3$\times$3 spiking convolutions. 

At a high level, input spikes are loaded by the spike streamer, which converts spike vectors into spike addresses. These addresses are sent to the clusters, where they are broadcast to the convolution units (CUs). The CUs use these addresses to fetch the corresponding weights and neuron states to perform accumulation in parallel. After all spikes within a timestep have been processed, the threshold unit in each cluster generates the output spikes.

\subsection{Spiking Convolution on Compute Clusters}
\label{sec:spikingconv}
Each compute cluster integrates 32 CUs, along with local neuron state memory and a threshold unit.

\subsubsection{Spiking convolution}
Convolutional Neural Networks (CNNs) commonly employ a sliding window, frame-based approach. For a $3 \times 3$ kernel $W$ and input $I$, the output at location $(x,y)$ is given by
\begin{equation}
    O[x,y] = \sum_{a=-1}^{1}\sum_{b=-1}^1I[x+a,y+b]W[a,b].
    \label{eq:conv}
\end{equation}
This formulation is output-centric, requiring all input elements contributing to a given output location to be fetched. In SNNs, where spikes are highly sparse, this approach leads to fetching many zeros, which makes processing inefficient.

An alternative approach, first proposed in \cite{tapiador-morales2019}, defines the convolution in an event-driven manner. For an incoming spike at location $(u,v)$, the spike $S[u,v]$ is included in the sum for neuron state $V[x,y]$ when
\begin{align}
    (u,v) = (x+a,y+b), \quad (a,b) \in \{-1,0,1\}^2,
\end{align}
which can be rewritten as
\begin{align}
    x = u - a,\quad y = v - b.
    \label{eq:convcoords}
\end{align}

This leads to the following update rule:
\begin{align}
    \Delta V[u-a,v-b] \mathrel{{=}} W[a,b] \text{ for } (a,b) \in \{-1,0,1\}^2.
    \label{eq:convupdate}
\end{align}
This event-driven update rule produces the same results as the frame-based approach. However, since the update rule is event-driven, computation scales with the number of input spikes. 

\begin{figure}[t]
    \centering
    \includegraphics[width=\linewidth,trim=0 3 0 3,clip]{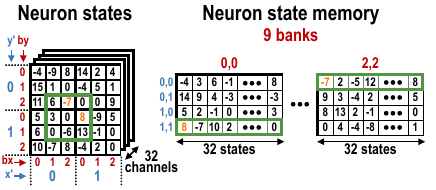}
    \caption{\textbf{Neuron state memory organization}. Neuron states are distributed across nine SRAM banks. The $(x,y)$ coordinate of a state is transformed into $(x',y',bx,by)$, as shown on the left. The bank indices $(bx,by)$ select one of the nine banks, while $(x',y')$ defines the word address within the bank. Each memory word stores 32 neuron states corresponding to 32 output channels.}
    \label{fig:neuronstatememory}
\end{figure}

\subsubsection{Neuron state memory}
The update rule in \eqref{eq:convupdate} allows for nine neuron updates in parallel. To fetch the nine neuron states in parallel, we extend the memory interlacing scheme used in \cite{sommer2022}. As shown in \autoref{fig:neuronstatememory}, neuron states are distributed across nine dual-ported SRAM banks, such that each $3 \times 3$ window of neuron states can be accessed concurrently. To further increase throughput, we exploit parallelism across output channels. Unlike in \cite{sommer2022}, each SRAM word stores 32 neuron states corresponding to 32 output channels. With nine SRAM banks, the architecture supports 288 concurrent neuron state reads and writes per cycle. Each SRAM bank has a capacity of \qty{16}{\kilo\byte}, resulting in a total neuron state memory of \qty{144}{\kilo\byte}.

\subsubsection{Convolution unit}
Mega employs nine compute clusters, each responsible for one of the nine kernel offsets in \eqref{eq:convupdate}, enabling fully parallel kernel updates. Within each cluster, 32 CUs operate in parallel to process 32 output channels, resulting in a total of 288 neuron updates per cycle.

Each CU implements a four-stage pipeline:
\begin{enumerate}
    \item \textbf{Address generation}: Compute the target neuron address from the input spike coordinates.
    \item \textbf{Neuron state fetch}: Read the 8-bit neuron state from the neuron state memory.
    \item \textbf{Neuron update}: Add the weight to the neuron state. Overflow is handled by clipping the neuron state.
    \item \textbf{Write-back}: Store the updated neuron state in the neuron state memory.
\end{enumerate}
Read-After-Write (RAW) hazards can occur when a neuron state is read before it has been updated or written back. These hazards are detected during the address calculation stage and are resolved by data forwarding, eliminating pipeline stalls.

\subsubsection{Threshold unit} The threshold unit applies leakage and thresholding. Mega employs leaky integrate-and-fire (LIF) neurons with linear decay. Thresholding is performed in a row-wise manner, activating three of the nine thresholding units concurrently. Output spikes are buffered before being written back to the shared memory.

The threshold unit also implements a four-stage pipeline:
\begin{enumerate}
    \item \textbf{Address calculation}: Neuron states are accessed sequentially. The address is generated using a counter.
    \item \textbf{Neuron state fetch}: Read 32 neuron states in parallel.
    \item \textbf{Leak and threshold}: Apply linear leakage to the signed 8-bit neuron states, followed by the threshold comparison. Neurons exceeding the threshold emit a spike.
    \item \textbf{Write-back}: Store the updated neuron states in the neuron state memory. Spiking neurons are reset to zero; otherwise, the leaked neuron state is written back.
\end{enumerate}

\subsection{SoC Integration}
To increase flexibility, Mega is integrated with the SNAX framework \cite{antonio2025}, which provides a RISC-V control system and a Tightly-Coupled Data Memory (TCDM). 

The RISC-V control system communicates with Mega through the Control and Status Registers (CSRs) interface of Mega, enabling the configuration of parameters such as neuron leakage and the threshold. The RISC-V controller also manages data transfers between Mega and the TCDM. The TCDM consists of 32 single-port SRAM banks of \qty{8}{\kilo\byte} each, totaling \qty{256}{\kilo\byte}. It stores spikes, neuron states, and weights before they are loaded into Mega's local memories. Mega can access the entire TCDM, providing flexibility to efficiently handle layers with either large neuron states or large weight kernels.

\subsection{Spike streamer}
The spike streamer converts dense binary spike maps into sparse spike addresses, enabling the spiking convolution explained in Section \ref{sec:spikingconv}. It fetches 96-bit spike vectors from the TCDM and finds spikes using a Leading Zero Counter (LZC), which is implemented efficiently as a tree structure. After a spike is extracted, the corresponding bit is cleared, allowing the LZC to locate the next spike. Once all spikes in a vector have been processed, the spike streamer proceeds to the next vector.

To avoid stalls when fetching the next spike vector, we prefetch it while the current vector is being processed. A second LZC operates on the prefetched vector, allowing immediate continuation without stalls.

The spike location $(x,y)$ is derived from the combination of the spike vector address and the LZC output. Due to the memory interlacing scheme used, the location needs to be converted to the $(x',y',bx,by)$ format. To avoid expensive divide-by-3 and modulo-3 operations, $y'$ and $by$ are generated using lightweight counters. However, since the $x$ coordinate depends directly on the LZC output, we cannot use counters to compute $x'$ and $bx$. Therefore, we employ two small lookup tables (LUTs) with 96 entries, which store the $x'$ and $bx$ corresponding to each LZC output.

\section{Measurement Results}

\begin{figure}[t]
    \centering
    \includegraphics[width=\linewidth]{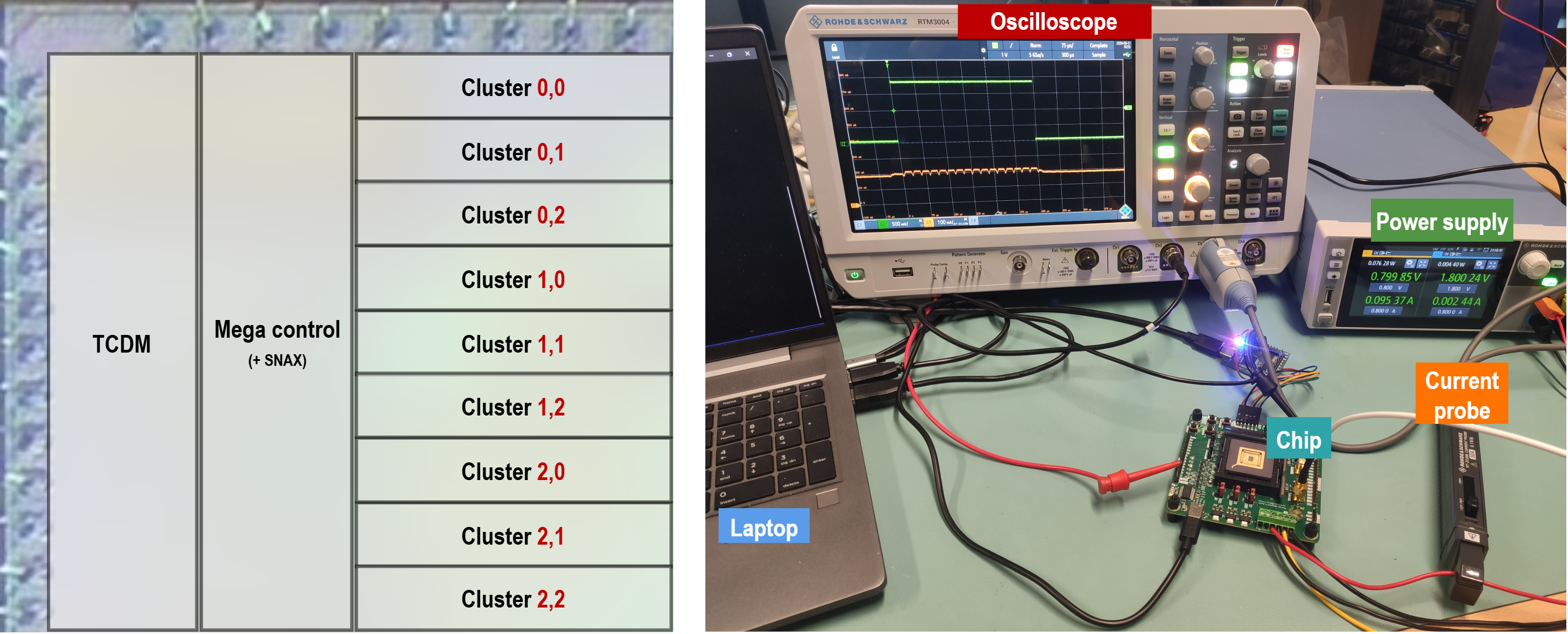}
    \caption{Chip micrograph and measurement setup}
    \label{fig:experimentalsetup}
\end{figure}

\input{chiptable}

\input{comptable}

\begin{figure}[t]
    \centering
    \includegraphics[width=\linewidth]{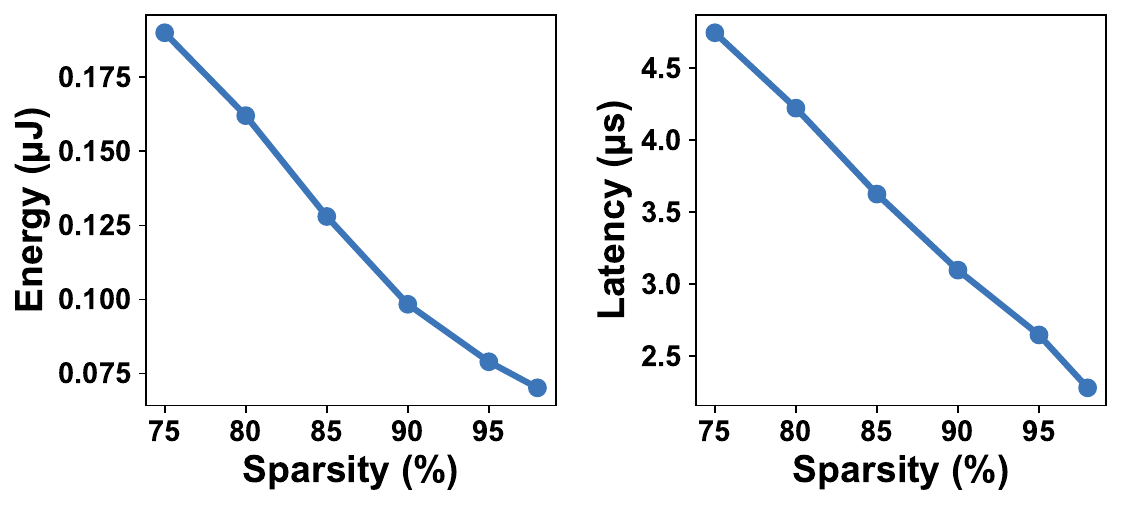}
    \caption{Energy and latency versus input sparsities for a $96 \times 50$ spike map. }
    \label{fig:sparsityplot}
\end{figure}

\begin{figure}[t]
    \centering
    \includegraphics[width=\linewidth]{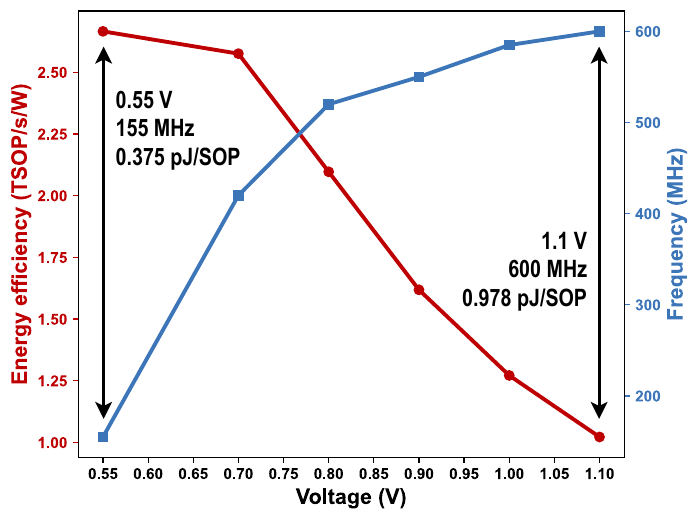}
    \caption{Energy efficiency and performance tradeoff.}
    \label{fig:voltageplot}
\end{figure}

\begin{figure}[t]
    \centering
    \includegraphics[width=\linewidth,trim=0 2 0 0,clip]{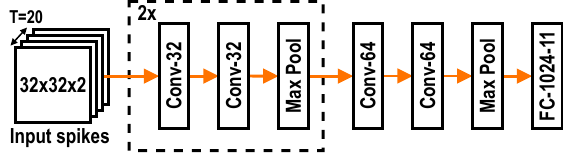}
    \caption{SNN architecture used for the DVS gestures task.}
    \vspace{-0.5em}
    \label{fig:convsnn}
\end{figure}

Mega is fabricated as part of a larger SoC tapeout in \mbox{GlobalFoundries} \qty{22}{\nano\meter} FDSOI. It occupies \qty{1.12}{\milli\meter\squared}, which is \qty{23.0}{\percent} of the SoC area. The chip micrograph and measurement setup are shown in \autoref{fig:experimentalsetup}. The chip summary is given in \autoref{tab:chiptable}.

As Mega is integrated within a larger SoC sharing a single power domain, its individual contributions to leakage and dynamic power must be determined. Leakage power is found by measuring the total leakage and scaling it proportionally to the area occupied by Mega. Dynamic power is obtained by measuring the total chip power with and without activity on Mega, and taking the difference between the measurements.

\autoref{fig:sparsityplot} shows the energy and latency of Mega across different input sparsity levels. Both the energy and latency scale well with the input sparsity as a result of the event-driven spiking convolution. \autoref{fig:voltageplot} presents the energy efficiency and performance tradeoff at different voltages. All tested samples of the chip operate from \qty{0.55}{\volt} to \qty{1.1}{\volt}. At \qty{0.55}{\volt}, Mega runs at up to \qty{155}{\mega\hertz}, consuming just \qty{0.375}{\pjsop}.

We evaluate Mega on the IBM DVS gestures dataset \cite{amir2017}, a widely used benchmark for event-based vision. The implemented network, shown in \autoref{fig:convsnn}, achieves an accuracy of \qty{96.1}{\percent}. Each convolutional layer is deployed and evaluated individually on Mega using inputs with representative sparsity levels. Aggregating the per-layer measurements, Mega consumes \qty{174.9}{\ujinf}, demonstrating its suitability for energy-efficient edge vision applications.

\autoref{tab:comp} compares Mega with recent state-of-the-art neuromorphic accelerators. ReckOn \cite{frenkel2022} and ANP-I\cite{zhang2023} target (recurrent) fully connected networks, limiting their scalability to larger vision workloads. In contrast, Mega supports convolutional SNNs, enabling efficient execution of more complex and larger-scale networks. \cite{liu2024, sharma2025, fu2026} adopt Compute-In-Memory (CIM) to support convolutional SNNs. Despite not using CIM, Mega achieves the highest energy efficiency, improving on prior work by $4\times$.

\section{Conclusion}
This paper presented Mega, a digital convolutional SNN accelerator. It combines event-driven spiking convolution with low-overhead spike detection for efficient operation. To improve flexibility, Mega employs a unified memory for spikes, states, and weights. Implemented in \qty{22}{nm}, Mega consumes \qty{0.375}{\pjsop}, achieving state-of-the-art energy efficiency.

\section*{Acknowledgment}
This work is funded by EU’s Horizon Europe research and innovation programme under grant agreement No. 101070374. The authors thank Henk Corporaal for his valuable feedback.

\bibliographystyle{./IEEEtran.bst}
\bibliography{references}

\end{document}

%% file: chiptable.tex
\begin{table}[t]
\centering
\renewcommand{\arraystretch}{1.2}
\caption{Chip summary}
\label{tab:chiptable}
\begin{tabular}{|c|cc|}
\hline
\textbf{Technology}             & \multicolumn{2}{c|}{22 nm FDSOI CMOS}                   \\ \hline
\textbf{Area}                   & \multicolumn{2}{c|}{1.12 mm$^2$}                          \\ \hline
\textbf{Supply voltage}         & \multicolumn{2}{c|}{0.55 V - 1.1 V}                     \\ \hline
\textbf{Frequency}              & \multicolumn{2}{c|}{155 MHz - 600 MHz}                  \\ \hline
\textbf{Network type}           & \multicolumn{2}{c|}{Convolutional SNN}                           \\ \hline
\textbf{Weight/State precision} & \multicolumn{2}{c|}{INT4/INT8}                          \\ \hline
                                & \multicolumn{1}{c|}{0.55 V @ 155 MHz} & 1.1 V @ 600 MHz \\ \hline
\textbf{Power consumption}      & \multicolumn{1}{c|}{14.4 mW}          & 144.7 mW        \\ \hline
\textbf{Energy efficiency}      & \multicolumn{1}{c|}{0.375 pJ/SOP}     & 0.978 pJ/SOP    \\ \hline
\textbf{Throughput}             & \multicolumn{1}{c|}{38.4 GSOP/s}       & 148.7 GSOP/s            \\ \hline
\end{tabular}
\vspace{-1em}
\end{table}

%% file: comptable.tex
\begin{table*}[ht]
\centering
\renewcommand{\arraystretch}{1.2}
\caption{Comparison of neuromorphic accelerators}
\label{tab:comp}
\begin{tabular}{|c|c|c|c|c|c|c|}
\hline
\textbf{} &
  \begin{tabular}[c]{@{}c@{}}ReckOn \cite{frenkel2022}\\ ISSCC'22\end{tabular} &
  \begin{tabular}[c]{@{}c@{}}ANP-I \cite{zhang2023}\\ ISSCC'23\end{tabular} &
  \begin{tabular}[c]{@{}c@{}}Sparse-IMC \cite{liu2024}\\ ISSCC'24\end{tabular} &
  \begin{tabular}[c]{@{}c@{}}SpiDR \cite{sharma2025}\\ ESSERC'25\end{tabular} &
  \begin{tabular}[c]{@{}c@{}}SpikeRAM \cite{fu2026}\\ ISSCC'26\end{tabular} &
  \textbf{\begin{tabular}[c]{@{}c@{}}Mega\\ This work\end{tabular}} \\ \hline
\textbf{Technology}               & 28 nm           & 28 nm      & 22 nm        & 65 nm   &     65/40 nm     & \textbf{22 nm}        \\ \hline
\textbf{Voltage}                 & 0.5-0.8V        & 0.56-0.9V  & 0.55-0.9V    & 0.9-1.2V       & 0.8-1.1V  &  \textbf{0.55-1.1V}    \\ \hline
\textbf{Clock frequency}        & 13-115 MHz      & Async      & 51-280 MHz   & 50-150 MHz     & 50-210 MHz & \textbf{155-600 MHz}  \\ \hline
\textbf{Network type}             & SRNN            & FC SNN     & Conv SNN     & FC/Conv SNN    & FC/Conv SNN & \textbf{Conv SNN}     \\ \hline
\textbf{State resolution}  & INT16      & (N/A) & (N/A) & INT7, INT11, INT15 & INT16 & \textbf{INT8}    \\ \hline
\textbf{Weight resolution}  & INT8      & INT8 & INT4, INT8 & INT4, INT6, INT8 & INT8 & \textbf{INT4}    \\ \hline
\textbf{Neuron model}           & LIF             & LIF        & LIF          & IF, LIF, RMP   & LIF & \textbf{LIF}          \\ \hline
\textbf{DVS gestures Accuracy}                &     \qty{87.3}{\percent}            &  \qty{92.0}{\percent}          &     \qty{94.0}{\percent}         &      \qty{93.5}{\percent}          &  \qty{92.9}{\percent} & \textbf{\qty{96.1}{\percent}}             \\ \hline
\textbf{Energy Efficiency}       & 5.3-12.8 pJ/SOP & 1.5 pJ/SOP & 3.87 pJ/SOP  & 5 TOPS/W$^*$       & 11.67 pJ/SOP  & \textbf{0.375 pJ/SOP} \\ \hline
\multicolumn{7}{l}{\scriptsize *Measured at \qty{95}{\percent} sparsity, includes operations skipped due to zeros.} 
\end{tabular}%
\vspace{-1em}
\end{table*}